\documentclass[12pt]{iopart}

\expandafter\let\csname equation*\endcsname\relax
\expandafter\let\csname endequation*\endcsname\relax
\usepackage[utf8]{inputenc}                             
\usepackage[final,nopatch=footnote]{microtype}
\usepackage[T1]{fontenc}                                
\usepackage{hyperref}                                   
\usepackage{url}                                        
\usepackage{booktabs}                                   
\usepackage{amsfonts,amssymb,amsmath}                   
\usepackage{bm,graphicx}
\usepackage{microtype}                                  
\usepackage{setspace} 
\usepackage[colorinlistoftodos]{todonotes}                

\usepackage{etoolbox}
\usepackage[normalem]{ulem}

\providecolor{added}{rgb}{0,0,1}
\providecolor{deleted}{rgb}{1,0,0}

\makeatletter
\newrobustcmd{\fixappendix}{%
  \patchcmd{\l@section}{1.5em}{7em}{}{}%
  \patchcmd{\l@subsection}{2.3em}{7em}{}{}%
}
\makeatother

\def\bea{\begin{eqnarray}}
\def\eea{\end{eqnarray}}
\def\la{\langle}
\def\ra{\rangle}

\makeatletter
\newcommand{\reqnomode}{\tagsleft@true}
\newcommand{\leqnomode}{\tagsleft@false}
\makeatother

\makeatletter
\def\@mkboth#1#2{}
\newlength\appendixwidth
\preto\appendix{\addtocontents{toc}{\protect\patchl@section}}
\newcommand{\patchl@section}{%
  \settowidth{\appendixwidth}{\textbf{Appendix }}%
  \addtolength{\appendixwidth}{1.5em}%
  \patchcmd{\l@section}{1.5em}{\appendixwidth}{}{\ddt}%
}
\makeatother

\begin{document}

\title[]{From Attraction to Repulsion: Emergent Interactions in Harmonically Coupled Active Binary System}

\author{Ritwick Sarkar}
\address{S. N. Bose National Centre for Basic Sciences, Kolkata 700106, India}
\ead{ritwick.sarkar@bose.res.in}
\author{Sreya Chatterjee}
\address{Friedrich-Alexander-Universit\"{a}t Erlangen-N\"{u}rnberg, 91054 Germany}
\ead{sreya.chatterjee@fau.de}
\author{Urna Basu}
\address{S. N. Bose National Centre for Basic Sciences, Kolkata 700106, India}
\ead{urna@bose.res.in}

\begin{abstract}
We investigate the emergent interactions between two active Brownian particles coupled by an attractive harmonic potential and in contact with a thermal reservoir. By analyzing the stationary distribution of their separation, we demonstrate that the effective interaction can be either attractive or repulsive, depending on the interplay between activity, coupling strength, and temperature. Notably, we find that an effective short-range repulsion emerges in the strong and moderate-coupling regimes, when the temperature is below some threshold value, which we characterize analytically. In the strong-coupling regime, the repulsion emerges \textcolor{black}{solely} due to the difference in the self-propulsion speeds of the particles. We also compute the short-time position distribution of the centroid of the coupled particles, which shows strongly non-Gaussian fluctuations at low temperatures.

\end{abstract}

\maketitle

\makeatletter

\tableofcontents

\section{Introduction}

The theory of active matter has grabbed the attention of physicists for over a decade now. Active matter systems consist of individual `active agents', which consume energy from their environments at an individual level and perform persistent motion~\cite{romanczuk2012active,bechinger_active_matter,ramaswamy2010mechanics, PhysRevX.15.021050}. Examples of active motion are found both in laboratory-synthesised matter, such as \textcolor{black}{Janus beads~\cite{howse_motile_colloid,PhysRevE.88.032304,PhysRevLett.123.098001}, colloidal rollers~\cite{bricard2015emergent}, synthetic nanomotors~\cite{Gaspard01012019}, programmable robots~\cite{ PRXLife.2.033007,paramanick2024programming} and in nature, ranging from motility of biofilaments~\cite{KUMAR2013583a}, molecular motors~\cite{schaller2010polar} and bacteria~\cite{berg2004coli} to collective dynamics of animals~\cite{annurev-conmatphys-031113-133834, PhysRevLett.75.4326, TONER2005170, PhysRevLett.120.198101, PhysRevE.107.024411}.} The intrinsically nonequilibrium nature of active particle dynamics gives rise to collective behaviors that are significantly more complex than those observed in equilibrium systems. \textcolor{black}{One of the most striking features of the active matter systems is their tendency to form aggregates~\cite{aannurev-conmatphys-031214-014710} in the presence of attraction~\cite{science.1230020}, repulsion~\cite{PhysRevE.99.032605, PhysRevLett.121.098003, PhysRevLett.124.078001}, alignment~\cite{PhysRevE.104.024130}, and even in the absence of any underlying interactions~\cite{10.1063_1.5134455}.}

Theoretical attempts to understand the origin of the unusual behavior in active particle systems often focus on studying simple model systems comprising a few particles. \textcolor{black}{Recent studies about a pair of active Run-and-Tumble particles moving on a one-dimensional lattice with hardcore exclusion interactions have shown that persistence in active dynamics gives rise to an effective attraction between the particles~\cite{slowman2016jamming, Slowman_2017, Das_2020}. A similar emergent attraction is also observed for the more general case of many particle systems in arbitrary dimensions~\cite{Mallmin_2019,Metson_2020}. Even in the presence of recoil interaction, active particles are found to exhibit effective attractive behavior~\cite{PhysRevE.107.044134, Metson_2023}, which is exceptional for active particles, as a similar recoil interaction between passive particles always results in an effective repulsion. Such emergent attraction is also found where active particles interact through long-range attractive potentials~\cite{PhysRevE.104.044103}.}  On the other hand, it has recently been shown that the presence of diversity in self-propulsion speed can lead to an emergent repulsion among active particles~\cite{sarkar2024emergent}. However, a comprehensive understanding of such emergent repulsion is still lacking.

In this work, we present an extensive study of a binary system of active Brownian particles coupled via a harmonic potential and in contact with a thermal reservoir. The separation between the two particles eventually reaches a nonequilibrium stationary state, while the centroid undergoes an unbounded persistent motion. We show that, depending on the relative strength of the coupling and rotational diffusion constants of the two particles, three distinct regimes emerge. We analytically characterize the nonequilibrium stationary state by computing the radial and $x$-marginal distribution of the separation vector in the three regimes. Surprisingly, we find that, at sufficiently low temperature, despite the underlying attractive interaction, the particles effectively repel each other when the coupling strength is large compared to at least one rotational diffusion constant. We also analytically characterize the short-time distribution of the centroid, which shows strongly non-Gaussian fluctuations.

The paper is organised as follows. We introduce the model and briefly summarise our main results in the next section. Section~\ref{sec:dist} focuses on the NESS of the separation in the three regimes. The position fluctuations of the centroid are discussed in Sec.~\ref{sec:cm}. We conclude with some general remarks in Sec.~\ref{sec:concl}.

\section{Model and results}
We consider two harmonically coupled overdamped active Brownian particles moving on a $2d$ plane, in contact with a thermal bath of temperature $T$. Let ${\bm r}_i = (x_i,y_i)$ denote the position of the $i$-th particle with $i=1,2$, and $k$ denote the coupling strength. Moreover, let $v_i$ denote the self-propulsion speed of the $i$-th particle along its internal orientation $\hat{\bm n}_i = (\cos{\theta_i}, \sin{\theta_i})$. The internal orientations of both particles undergo rotational diffusion with diffusion coefficients $D_1$ and $D_2$, respectively.  The Langevin equations governing the time-evolution of the positions of the particles are given by,
\begin{align}
\dot{\bm{r}}_1 = -k(\bm{r}_1 - \bm{r}_2) + v_1 \hat{\bm n}_1 + \sqrt{2T}\, {\bm \xi}_1(t), \label{eq:model_r1} \\
\dot{\bm{r}}_2 = -k(\bm{r}_2 - \bm{r}_1) + v_2 \hat{\bm n}_2 + \sqrt{2T}\, {\bm \xi}_2(t), 
\label{eq:model_r2}  
\end{align} 
where $\{ {\bm \xi}_i =(\xi_i^x, \xi_i^y); i=1,2 \}$ denote independent white noises  with zero mean and correlations given by
\begin{align}
\la \xi_i^\alpha(t) \xi_j^\beta (t') \ra =\delta_{ij} \delta_{\alpha \beta}\delta{(t-t')}.
\end{align}
The internal orientations $\{\theta_i\}$ evolve according to,
\begin{align}
\dot{\theta_{i}}(t) &=\sqrt{2D_{i}}\, \eta_{i}(t).
\label{eq:model_phi} 
\end{align}
Here, $\{\eta_{i}(t)\}$ denote another set of independent white noises with zero mean and correlation $\la \eta_i(t) \eta_j(t') \ra = \delta_{ij}\delta(t-t')$. We assume that both the particles are at the origin at $t=0$ with random initial orientations, i.e., $x_i(0)=y_i(0)=0$ and $(\theta_1(0), \theta_2(0))$ are chosen independently from the uniform distribution in $[0,2\pi]$. Therefore, the components of the active noises $\hat{\bm n}_i$ have zero mean and the following autocorrelation~\cite{drabp,abp_2D},
\begin{align}
\la \cos \theta_i(t) \cos \theta_j(t') \ra &= \la \sin \theta_i(t) \sin \theta_j(t') \ra =\delta_{ij}\frac{1}{2}e^{-D_i|t-t'|}.
\label{active_noise_correlation}
\end{align}

To understand the behavior of this two-particle system, it is convenient to rewrite Eqs.\eref{eq:model_r1}-\eref{eq:model_r2} in terms of the relative coordinate ${\bm r}={\bm r}_1 - {\bm r}_2$ and the centroid position ${\bm R}= ({\bm r}_1+{\bm r}_2)/2$. From Eqs.~\eref{eq:model_r1}-\eref{eq:model_r2} it is straightforward to see that the relative coordinate ${\bm r}$ follows a Langevin equation,
\begin{align}
\dot {\bm r}(t) &= - 2k {\bm r}(t) + v_1 \hat {\bm n}_1(t) - v_2 \hat {\bm n}_2(t)+\sqrt{4T}{\bm \xi}(t), ~~ \text{with} ~~ {\bm \xi}(t)=\frac{{\bm \xi}_1(t) - {\bm \xi}_2(t)}{\sqrt{2}}.\label{eom_rel_dist}
\end{align}
On the other hand, the centroid position evolves according to,
\begin{align}
 \dot {\bm R}(t) &= \frac{1}{2}\left[v_1 \hat {\bm n}_1(t) + v_2 \hat {\bm n}_2(t)\right]+\sqrt{T}{\bm \zeta}(t),~~ \text{with} ~~{\bm \zeta}(t)=\frac{{\bm \xi}_1(t) + {\bm \xi}_2(t)}{\sqrt{2}}. \label{eom_cm}   
\end{align}
Clearly, ${\bm \xi}$ and ${\bm \zeta}$ are two independent white noises with zero mean and correlators $\la \xi^\alpha(t) \xi^\beta(t') \ra = \delta_{\alpha,\beta} \delta(t-t')$, $\la \zeta^\alpha(t) \zeta^\beta(t') \ra = \delta_{\alpha,\beta} \delta(t-t')$ and $\la \xi^\alpha(t) \zeta^\beta(t')\ra=0$.

Equation~\eref{eom_rel_dist} implies that the motion of the relative coordinate resembles that of a particle in a harmonic potential of strength $2k$ in contact with a thermal reservoir at temperature $2T$, driven by the active noise $v_1 \hat {\bm n}_1 - v_2 \hat {\bm n}_2$. This motion is characterized by three different time-scales---the relaxation time-scale $(2k)^{-1}$ associated with the harmonic trap and the active time-scales $D_1^{-1}$ and $D_2^{-1}$ associated with the rotational diffusion of the internal orientations of the two particles. The relative coordinate $\bm{r}$ is expected to reach a stationary state at times much larger than all three time-scales.

On the other hand, Eq.~\eref{eom_cm} implies that the centroid undergoes an unbounded motion driven by a combination of active noises and thermal noise. In the absence of the thermal bath, the motion of the centroid is bounded in the annular region $C_\mathrm{min}\leq R \equiv |{\bm R}(t)|\leq C_\mathrm{max}$ with,
\begin{align}
C_\mathrm{min}=\frac{|v_1-v_2|t}{2}\quad \text{and}\quad C_\mathrm{max}= \frac{(v_1+v_2)t}{2}.\label{bound_cm}
\end{align}
\begin{figure}[t]
    \centering
    \includegraphics[width=0.9\linewidth]{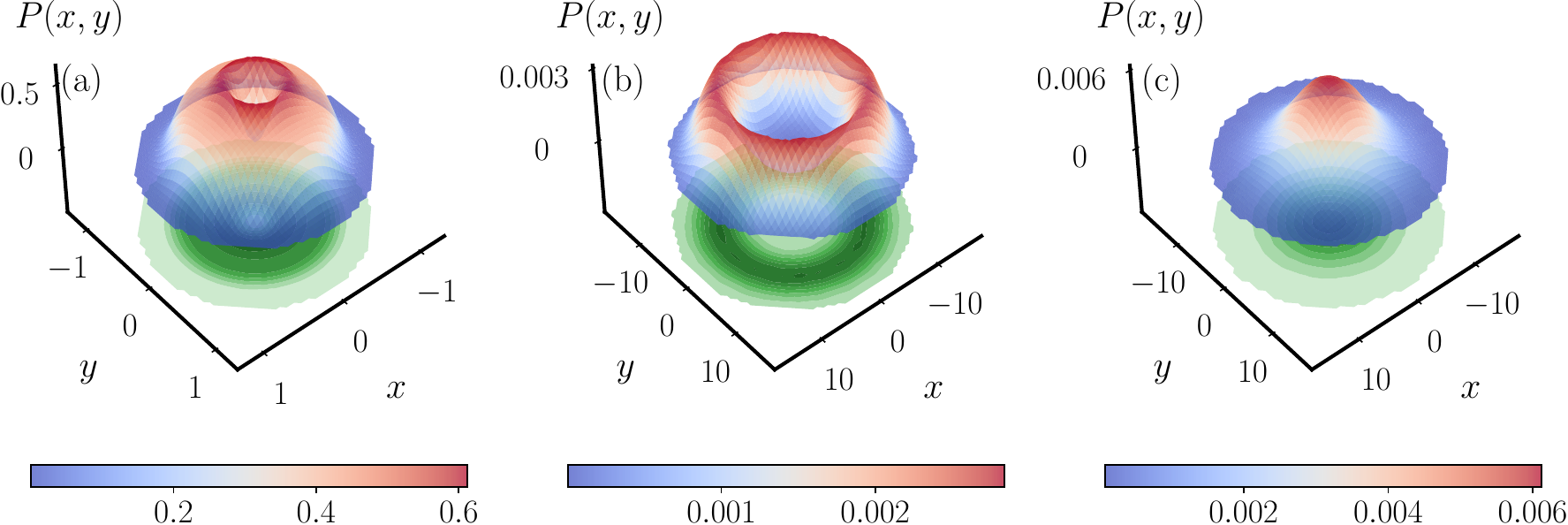}
    \caption{Stationary distribution $P\left((x,y)\right)$ of the relative coordinate $\bm{r}=(x,y)$, obtained from numerical simulations, in the strong-coupling regime (a), moderate-coupling regime (b), and weak-coupling regime (c). The projection on the $x-y$ plane is shown by the green contour plot. For (a), the parameters are $D_1=D_2=0.01, k=4, T=0.05, v_1=2$ and $v_2=4$. The self-propulsion velocities are fixed at $v_1=4, v_2=1$ for the moderate and weak-coupling regimes. The remaining parameters for (b) are $D_1=0.001, D_2=10, k=0.2, T=1$, and for (c) $D_1=10, D_2=20, k=0.02, T=0.1$.}
    \label{fig:surface}
\end{figure}

In this work, we present a comprehensive study of the stationary distribution of the relative coordinate $P({\bm r})$  and show that it has strikingly different features depending on the relative strengths of the different time-scales. In particular, we analytically compute the stationary distribution $P({\bm r})$ in the three limiting scenarios ---
\begin{itemize}
    \item Strong-coupling regime: $k\gg (D_1,D_2)$,
    \item Moderate-coupling regime: $D_1\ll k \ll D_2$, and
    \item Weak-coupling regime: $ (D_1, D_2)\gg k$.
\end{itemize}
Figure~\ref{fig:surface} shows plots of the typical stationary distribution $P\left({\bm r}=(x,y)\right)$ in these three different regimes.

The strong-coupling regime is characterized by the emergence of an unusual effective short-range repulsion between the particles for $v_1 \ne v_2$.  This was first shown in Ref~\cite{sarkar2024emergent} where the stationary distribution of the relative coordinate was computed in the absence of thermal noise. Here we analytically compute the marginal distributions of the radial distance $r$ and its $x$-component for $T>0$ and show that the repulsion survives for finite temperatures. To quantify this emergent repulsion, we define an effective potential which remains repulsive below a threshold temperature. In the moderate-coupling regime, the time evolution of the relative coordinate is equivalent to the motion of a single ABP in a harmonic trap with an effective temperature. The effective repulsion is also present in this regime, that vanishes with increasing $T_\mathrm{eff}$. 

The emergent repulsion disappears in the weak-coupling regime, where both the particles behave like passive ones and the typical distribution of the components of the separation vector $\bm {r}$ becomes Gaussian.  Finally, we also investigate the motion of the centroid. At short-times, i.e., for $t \ll ((2k)^{-1}, D_1^{-1}, D_2^{-1})$ it shows strongly non-Gaussian fluctuations, which we characterize analytically. 

\section{Stationary distribution of the separation} \label{sec:dist}

In this section, we focus on the stationary fluctuations of the separation variable ${\bm r}(t)$. The probability  that the particles are separated by a distance ${\bm r}$ at time $t$ can be formally expressed as $P({\bm r},t) = \int d\theta_1 d \theta_2 {\cal P}({\bm r}, \theta_1, \theta_2,t)$ where the joint distribution ${\cal P}({\bm r}, \theta_1, \theta_2,t)$ satisfies the Fokker-Planck equation,

\begin{align}
    \frac{\partial }{\partial t}{\cal P}({\bm r},\theta_1,\theta_2,t)=\nabla \cdot\left[(2k {\bm r}- v_1 \hat {\bm n}_1 + v_2 \hat {\bm n}_2){\cal P}\right]+4T\nabla^2 {\cal P} + D_1\frac{\partial^2 {\cal P}}{\partial \theta_1^2}+D_2\frac{\partial^2 {\cal P}}{\partial \theta_2^2}.\label{fp_eq}
\end{align}
Equation~\eref{fp_eq} is hard to solve even in the steady state. Instead, we focus on the Langevin Eq.~\eqref{eom_rel_dist} and adopt a trajectory-based approach to explore the statistical fluctuations of the separation ${\bm r}$. To this end, we write the formal solution of Eq.~\eref{eom_rel_dist},
\begin{align}
    {\bm r}(t)&=\int_0^t ds \, e^{-2 k (t-s)}\left[ v_1 \hat {\bm n}_1(s) - v_2 \hat {\bm n}_2(s)+\sqrt{4T}{\bm \xi}(s) \right],\label{x_comp_r}
\end{align}

To get a preliminary idea about the fluctuation of $r(t)$, we first calculate the mean and variance of the separation ${\bm r}(t)$.
Since the initial ${\theta_i}$ is chose from the uniform distribution in $[0,2\pi]$,  we have $\la \cos \theta_i(t) \ra =0 = \la \sin \theta_i(t) \ra$ for all $t$. Moreover, we have assumed both the particles start from the origin, ${\bm r}(0)=0$ and hence, both $\la x(t) \ra$ and $\la y(t) \ra$ vanish at all times. 

The second moment of the $x(t)$ and $y(t)$ can also be computed from Eq.~\eref{x_comp_r} as,
\begin{align}
\la x^2(t) \ra &=\int_0^t ds \int_0^t ds' e^{-2k(2 t-s-s')}\Big[v_1^2\la \cos\theta_1(s)\cos{\theta_1(s')} \ra+v_2^2\la \cos\theta_2(s)\cos{\theta_2(s')} \ra\cr
&\qquad  +4T\la \xi_x(s)\xi_x(s') \ra\Big], \label{eq:x2_t} \\
\la y^2(t) \ra &=\int_0^t ds \int_0^t ds' e^{-2k(2 t-s-s')}\Big[v_1^2\la \sin\theta_1(s)\sin{\theta_1(s')} \ra+v_2^2\la \sin\theta_2(s)\sin{\theta_2(s')} \ra\cr
& \qquad +4T\la \xi_y(s)\xi_y(s') \ra\Big]. \label{eq:y2_t}
\end{align}
Using the active noise correlations in Eq.~\eref{active_noise_correlation} and evaluating the integrals, we have,
\begin{align}
\la x^2(t) \ra = \la y^2(t) \ra &=  v_1^2\mathcal{G}(k,D_1,t)+v_2^2\mathcal{G}(k,D_2,t)+ \frac{T}{k}\big(1-e^{-4kt}\big). \label{2nd_moment_of_x}
\end{align}
where $\mathcal{G}(k,D,t)$ is given by, 
\begin{align}
    \mathcal{G}(k,D,t)=\frac{1-e^{-4kt}}{4 k(D+2k)}+\frac{e^{- 2 k t}(e^{-Dt}-e^{-2 k t})}{D^2-4 k^2}.
\end{align}

\begin{figure}[t]
    \centering
    \includegraphics[width=0.45\linewidth]{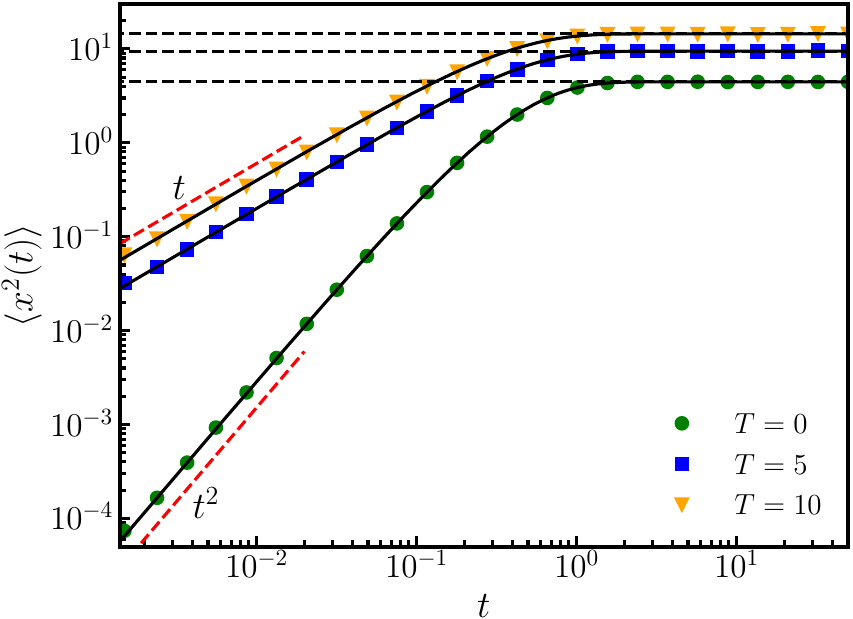}
    \caption{ Plot of $\la x^2(t) \ra$, the second moment of the $x$-component of the radial separation ${\bm r}$,  as a function of time for different values of the temperature $T$.  The symbols indicate the data obtained from the numerical simulations, whereas the black solid lines correspond to the theoretical prediction Eq.~\eqref{2nd_moment_of_x}. The dashed lines indicate the stationary values \eqref{stationary_x2}. Here we have taken $k=1$, $D_1=1$, $D_2=4$, $v_1=7$, and $v_2=3$. }
    \label{fig:moment_r}
\end{figure}

At short-times, i.e., for $t \ll (D_1^{-1}, D_2^{-1}, k^{-1})$, we have, from Eq.~\eqref{2nd_moment_of_x},
\begin{align}
    \la x^2(t) \ra&\simeq 4 T t-\Big( 8 kT-\frac{v_1^2+v_2^2}{2}\Big)t^2
    +\mathcal{O}[t^3].
\end{align}
Clearly, in the absence of a thermal bath, $\la x^2(t) \ra$ grows ballistically at short-times. However, for $T>0$, the short-time behaviour becomes diffusive. At large times, i.e., $t \gg (D_1^{-1}, D_2^{-1}, k^{-1})$, $\la x^2(t) \ra$ reaches a stationary value,
\begin{align}
    \la x^2 \ra=\frac{T}{k}+ \frac{v_{1}^2}{4k(D_{1}+2k)}+\frac{v_{2}^2}{4k(D_{2}+2k)}.\label{stationary_x2}
\end{align}
Figure~\ref{fig:moment_r} illustrates the behaviour of $\la x^2(t) \ra$ for different values of $T$.

In the following, we investigate the stationary state fluctuations of the relative coordinate ${\bm r}$ in the three regimes separately.

\subsection{Strong-coupling regime}

The strong-coupling regime emerges when the coupling strength $k$ is much larger than the rotational diffusion constants of the particles $D_1, D_2$. Consequently, the typical relaxation time of ${\bm r}$ in the trap $k^{-1}$ is much smaller than the time $D_i^{-1}$ taken by the orientations $\hat {\bm n}_i$  of the particles to change appreciably. As shown below, this allows one to obtain the stationary distribution exactly.

The radial and $x$-marginal distributions in the strong-coupling regime in the absence of temperature have recently been computed analytically, and it has been shown that heterogeneity in self-propulsion speed leads to the emergence of a short-range repulsion among the two particles~\cite{sarkar2024emergent}. More specifically, it was found that the distance $r = |{\bm r}|$ between the two particles must be bound in the regime  $R_\mathrm{min} \le r \le  R_\mathrm{max}$, with, 
\begin{align}
R_\mathrm{min}=\frac{|v_1-v_2|}{2k} \text{~and~}  R_\mathrm{max}=\frac{v_1+v_2}{2k}.
\end{align}
The radial distribution, in this case, follows a scaling form, 
\begin{align}
    P(r) = \frac 1{R_\text{max}} {\cal R} \left(\frac r{R_\text{max}} \right), 
\end{align}  
with the scaling function,
\begin{align}
 {\cal R}(u) =\frac{u}{\pi \sqrt{(1-u^2)(u^2- R_\text{min}^2/R_\text{max}^2)}} ~~ \text{for}~~ R_\text{min}/R_\text{max} \le u \le 1.
\end{align}
In the following, we investigate the stationary distribution in the presence of thermal noise.

As mentioned before, the strong-coupling regime is characterized by a time-scale separation between the relaxation of ${\bm r}$ in the potential and the evolution of orientations. Moreover, Eq~\eqref{eq:model_phi} implies that the orientations $(\theta_1, \theta_2)$ eventually relax to independent uniform distributions in $[0,2\pi]$. Hence, to the leading order in $D_i^{-1}$, the stationary distribution of the separation vector must follow,
\begin{align}
P({\bm r}) = \int_0^{2 \pi} \frac{d \theta_1}{2 \pi} \int_0^{2 \pi} \frac{d \theta_2}{2 \pi} ~ {\cal P}({\bm r}|\theta_1, \theta_2),\label{eq:dist_int}
\end{align}
where ${\cal P}({\bm r}|\theta_1, \theta_2)$ denotes the stationary distribution of the separation ${\bm r}$ for fixed $(\theta_1, \theta_2)$. The conditional distribution  ${\cal P}({\bm r}|\theta_1, \theta_2)$ can be computed from the Langevin equation Eq.~\eqref{eom_rel_dist} in a straightforward manner by taking the limit $t \gg k^{-1}$. In this limit, Eq.~\eqref{eom_rel_dist} can be recast as, 
\begin{align}
    {\bm r}=\frac{1}{2 k}\left(v_1\hat{\bm n}_1-v_2\hat{\bm n}_2\right)+{\bm w}.\label{radius}
\end{align}
where ${\bm w}= (w_x, w_y)$ is a Gaussian noise with independent components $w_x$ and $w_y$, distributed according to,
\begin{align}
    G(w_\alpha)=\sqrt{\frac{k}{2 \pi T}}\exp{\left(-\frac{k \, w_\alpha^2}{2 T}\right)},~~\mathrm{with}~~\alpha=x,y.\label{thermal_dist}
\end{align}
Using Eqs.~\eqref{eq:dist_int} and \eqref{radius}, we can now obtain,
\begin{align}
{\cal P}({\bm r}|\theta_1, \theta_2) = \int dw_x \int dw_y ~ \delta(\bm{r-q-w}) G(w_x) G(w_y), \label{eq:Pr_th_int}
\end{align}
where, for notational convenience, we have introduced ${\bm q} = \frac{1}{2 k}\left(v_1\hat{\bm n}_1-v_2\hat{\bm n}_2\right)$. Obviously,  
\begin{align}
 q_x = \frac 1{2k}(v_1 \cos \theta_1 - v_2 \cos \theta_2),~\quad \text{and},\quad q_y = \frac 1{2k}(v_1 \sin \theta_1 - v_2 \sin \theta_2).   
\end{align}
The stationary distribution of the separation in the strong-coupling regime can be computed using Eqs.~\eref{radius}-\eref{eq:Pr_th_int} along with Eq.~\eqref{eq:dist_int}. In the following, we compute the radial and $x$-component marginal distributions explicitly.

\subsubsection{Radial distribution:}\label{strong_couple_pr}

The radial distribution $P(r)$ denotes the probability that the radial distance between the two particles is $r$ in the stationary state and is normalized as $\int dr P(r)=1$. To obtain this distribution, it is convenient to first compute the probability distribution $\mathbb{P}(r^2)$ of $r^2 = x^2 +y^2$. The radial distribution is then obtained as  $ P(r) = 2r \mathbb{P}(r^2)$. To compute $\mathbb{P}(r^2)$, we start from Eq.~\eref{eq:Pr_th_int} and formally write, 
\begin{align}
   {\mathbb P}(z=r^2|\theta_1,\theta_2)=\int_{-\infty}^{\infty} d w_x \int_{-\infty}^{\infty} d w_y G(w_x)G(w_y)\delta\left[z-(q^2+w^2+ 2q_xw_x+2 q_yw_y)\right],
\end{align}
where, $q^2 = [v_1^2 +v_2^2 - 2 v_1 v_2 \cos (\theta_1 - \theta_2)]/(4k^2)$ and $w^2= w_x^2 + w_y^2$. To proceed further, we take the Laplace transform of the above equation with respect to $z$ and evaluate the Gaussian integrals over $w_x$ and $w_y$, which leads to,
\begin{align}
   \tilde  {\mathbb P}(s|\theta_1,\theta_2) &= \int_0^\infty dz\, e^{-sz} {\mathbb P}(z|\theta_1,\theta_2) \cr
   &= \frac{k}{k+2 s T}\exp{\left(-\frac{s(v_1^2+v_2^2)}{4 k (k+2 sT)}\right)}\exp{\left(\frac{s v_1 v_2\cos{(\theta_1-\theta_2)}}{2 k(k+2 s T)}\right)}.
\end{align}
\begin{figure}[t]
    \centering
    \includegraphics[width=0.7\linewidth]{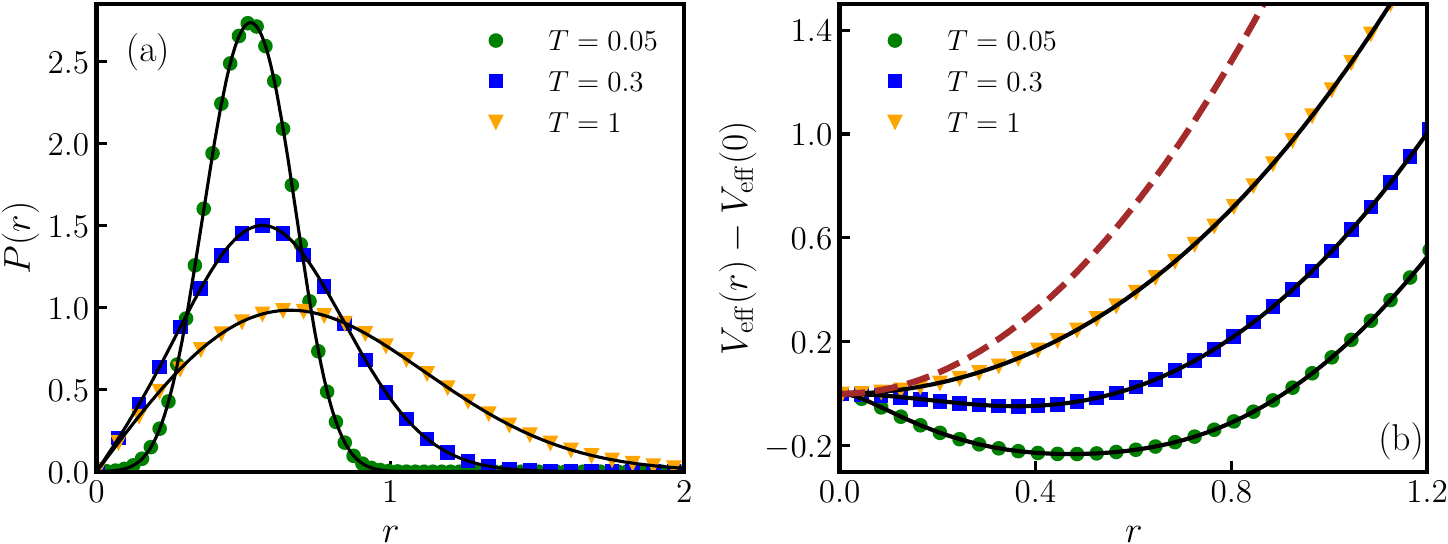}
    \caption{Strong-coupling regime: Plot of the stationary radial distribution  $P(r)$ (a) and the corresponding effective potentials (b), for different values of $T$, with $v_1=4$, $v_2=1$, $D_1=D_2=0.01$, and $k=4$. The black solid lines in (a) and (b) correspond to Eqs.~\eqref{radial_dist_strong} and \eqref{eff_pot2} respectively. The brown dashed line in (b) corresponds to the underlying harmonic potential $k r^2/2$ with $k=4$.}
    \label{fig:strong_coupling_eff_pot}
\end{figure}
Integrating over $\theta_1$ and $\theta_2$ [see Eq~\eqref{eq:dist_int}], we get the Laplace transform of ${\mathbb P}(z)$ as,
\begin{align}
\tilde {\mathbb P}(s) = \frac k{k+ 2 s T} \exp{\left[-\frac{s(v_1^2+v_2^2)}{4 k(k+ 2 s T)}\right]} I_0 \left(\frac{ v_1 v_2 s}{2k(k+ 2 s T)} \right),\label{laplace_density}  
\end{align}
where $I_0(u)$ denotes the zeroth-order modified Bessel function of the first kind. It is hard to invert the Laplace transform and get a closed-form expression for ${\mathbb P}(r^2)$. Instead, we expand the exponential and $I_0(u)$ and compute the inverse Laplace transform term by term, which leads to,
\begin{align}
 {\mathbb P}(z)=  \sum_{n=0}^\infty\sum_{m=0}^\infty (-1)^n \frac{(v_1^2+v_2^2)^n(v_1 v_2)^{2 m}4k^2}{n! (m!)^2(8 kT)^{2m+n+1}}{}_1F_1\left(2m+n+1,1,-\frac{k z}{2 T}\right),\label{pz_app}  
\end{align}
where ${}_1F_1(a;b;z)$ is the confluent hypergeometric function of the first kind or Kummer's function \cite{DLMF}. The sum over $m$ can be performed exactly and yields the radial distribution,
\begin{align}
    P(r)&= 2r {\mathbb P}(r^2) \cr 
    &= \frac{kr}{T}\sum_{\ell=0}^\infty\frac{(-(v_1^2+v_2^2))^\ell}{\ell!\, (8 kT)^\ell}\,  {}_2F_1\left[\frac{1-\ell}2, - \frac \ell 2,1, \frac{4 v_1^2 v_2^2}{(v_1^2+v_2^2)^2}\right] {}_1F_1\left[\ell+1,1,-\frac{k r^2}{2 T}\right]. \label{radial_dist_strong}
\end{align}
Here ${}_2F_1(a,b;c;z)$ in Eq.~\eqref{radial_dist_strong} is Gauss's hypergeometric function \cite{DLMF}. Although the sum over $\ell$ cannot be evaluated analytically, the radial distribution can be obtained to arbitrary accuracy by evaluating the sum numerically. Figure~\ref{fig:strong_coupling_eff_pot}(a) compares this theoretical prediction with the radial distribution $P(r)$ obtained from numerical simulations for different values of $T$.  It is evident that $P(r)$ is a single-peaked distribution that becomes broader as the temperature increases.

To understand the emergent interaction between the two particles quantitatively, it is useful to define an `effective potential'~\cite{slowman2016jamming, Seifert_2012, Marconi01092016, Pototsky_2012,to2014boltzmann, Ferrer_2024}. This is usually done by recasting the radial distribution in the form,
\begin{align}
    P(r)=A \,r \exp{\left[-\frac{V_\mathrm{eff}(r)}{T}\right]}, \label{eff_pot}
\end{align}
where $A$ is the normalization constant. Using the above definition,  the effective potential can be extracted from Eq.~\eref{radial_dist_strong}, or equivalently, the $P(r)$ measured from numerical simulations, as
\begin{align}
    V_\mathrm{eff}(r)-V_\mathrm{eff}(0)=-T \log{\left(\frac{P(r)}{r}\right)}+ T\left[\log{\left(\frac{P(r)}{r}\right)}\Bigg|_{r=0} \right ].\label{eff_pot2}
\end{align}
Figure~\ref{fig:strong_coupling_eff_pot}(b) shows a plot of the effective potential for different values of $T$. It is evident from this plot that for low temperatures, the effective potential $V_\text{eff}(r)-V_\text{eff}(0)$ has a minimum at some $r^*>0$, indicating that the typical distance between the two particles is finite. This, in turn, implies that, at low temperatures, there is an emergent repulsive interaction between the two particles. The repulsion disappears, evidenced by $r^*$ approaching zero, as the temperature is increased beyond a threshold value $T^*$. 
\begin{figure}[t]
    \centering
    \includegraphics[width=0.7\linewidth]{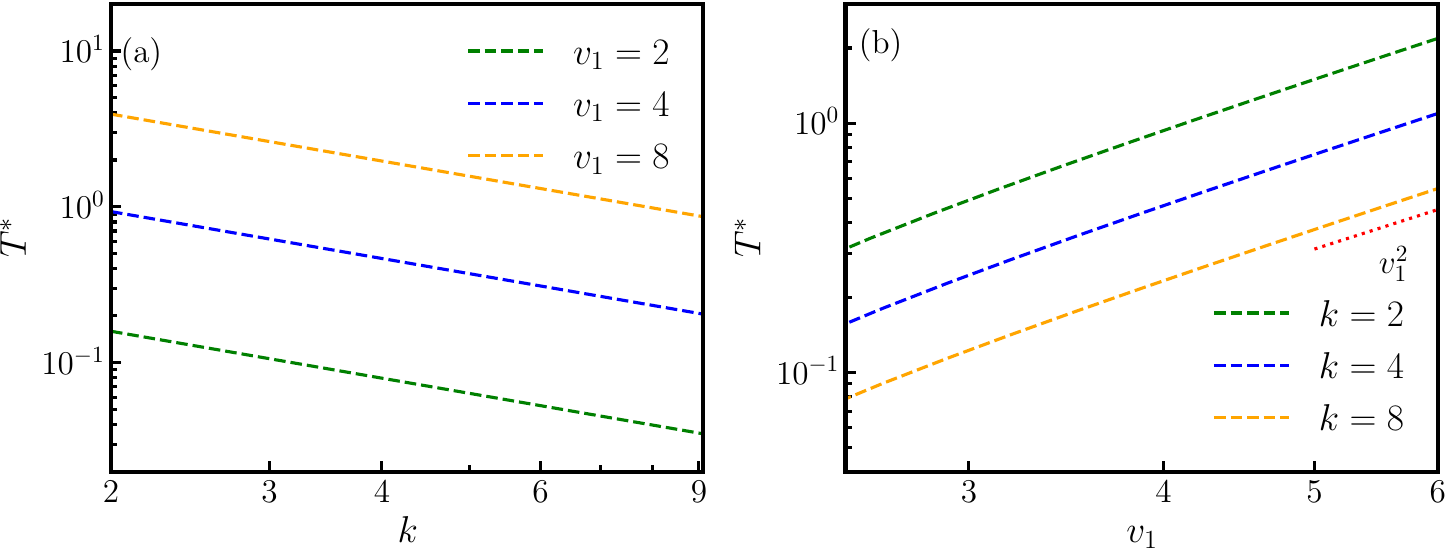}
    \caption{Strong-coupling regime: Plot of threshold temperature $T^*$ obtained by numerically solving Eq.~\eqref{b_eq} as a function of (a) coupling strength $k$ for different values of $v_1$, and (b) self-propulsion velocity $v_1$ for different values of $k$ with fixed $v_2=1$.}
    \label{fig:threshold_temp}
\end{figure}

To understand how the threshold temperature depends on $(v_1,v_2)$ and $k$, we look at the behaviour of the effective potential near the origin $r=0$. It is clear from Eq.~\eqref{radial_dist_strong} that $P(r)/r$ is a function of $r^2$, and hence, we must have, near $r=0$,
\begin{align}
V_\mathrm{eff}(r)-V_\mathrm{eff}(0) = B r^2 +\mathcal{O}[r^4],    
\end{align}
where $B$ depends on $v_1, v_2, k$ and $T$. Clearly, if $B<0$ 
then the effective potential is repulsive near $r=0$ and   $B>0$ indicates an effective attraction. The coefficient $B$ can be explicitly computed from the analytical expression of $P(r)$. To this end, it is convenient to start from Eq.~\eqref{pz_app}, and expand the right hand side around $r=0$ to get,  
\begin{align}
    \frac{P(r)}{r}&=\sum_{n=0}^\infty\sum_{m=0}^\infty (-1)^n \frac{(v_1^2+v_2^2)^n(v_1 v_2)^{2 m}8k^2}{n! (m!)^2(8 kT)^{2m+n+1}}\left(1-\frac{k r^2 (2 m+n+1)}{2 T}+\mathcal{O}[r^4]\right),\cr
    &\simeq\frac{k}{T}e^{-\frac{v_1^2+v_2^2}{8kT}}I_0\left(\frac{v_1 v_2}{4 kT}\right)\left[1-\frac{r^2}{16 T^2}\left(8kT-v_1^2-v_2^2+2v_1v_2\frac{I_1\left(\frac{v_1 v_2}{4 kT}\right)}{I_0\left(\frac{v_1 v_2}{4 kT}\right)}\right) + \mathcal{O}[r^4] \right].\cr\label{eq:rad_12}
\end{align}
Computing the effective potential from the above equation, we get,
\begin{align}
B = 8kT-v^2_1-v^2_2+2v_1v_2\frac{I_1\left(\frac{v_1 v_2}{4 kT}\right)}{I_0\left(\frac{v_1 v_2}{4 kT}\right)}.\label{b_eq}
\end{align}
The threshold temperature satisfies $B(T^*)=0$. Although it is hard to solve this equation analytically, it can be solved numerically to obtain the threshold temperature $T^*$. Figure~\ref{fig:threshold_temp}(a) and (b) show the variation of $T^*$ with $k$ and $v_1$, respectively. Clearly, the threshold temperature increases, i.e., the repulsion survives for larger thermal noise when $v_1 - v_2$ is larger.

\subsubsection{Marginal $x$-distribution:} In this section, we compute the marginal distribution $P(x)$ of the $x$-component of the separation vector in the strong-coupling regime. In the absence of the thermal noise, $P(x)$ follows a scaling form~\cite{sarkar2024emergent},
 \begin{align}
    P(x) = \frac{1}{R_{\text{max}}} {\cal F} \left (\frac{x}{R_{\text{max}}} \right),\label{strong_coupling_temp_zero}
 \end{align}
with the scaling function 
\begin{align}
{\cal F}(u) = \begin{cases}
\displaystyle \frac{2i}{\pi^2 \sqrt{R^2_{\text{min}}/R^2_{\text{max}} -u^2}} \Bigg[K\bigg (\frac{1 - u^2}{R^2_{\text{min}}/R^2_{\text{max}} - u^2}\bigg)  \cr 
\displaystyle \, - \sqrt{\frac{R^2_{\text{min}}/R^2_{\text{max}} -u^2}{1 -u^2}} K\bigg (\frac{R^2_{\text{min}}/R^2_{\text{max}} - u^2}{1 - u^2} \bigg) \Bigg] ~\text{for} ~~ |u| < R_{\text{min}}/R_{\text{max}}, ~~\cr 
\displaystyle \frac{2}{\pi^2 \sqrt{u^2 - R^2_{\text{min}}/R^2_{\text{max}}}}  K\Bigg [\frac{1 - u^2}{R^2_{\text{min}}/R^2_{\text{max}} - u^2} \Bigg] ~\text{for} ~~ R_{\text{min}}/R_{\text{max}} < |u| < 1,
\end{cases}
\end{align}
where $K(u)$ denotes the complete Elliptic integral of the first kind \cite{DLMF}. The emergent repulsion is apparent from the double-peaked nature of the distribution. For non-zero temperature, also such a double-peaked nature of $P(x)$ is expected for $T<T^*$.

\begin{figure}[t]
    \centering
    \includegraphics[width=0.7\linewidth]{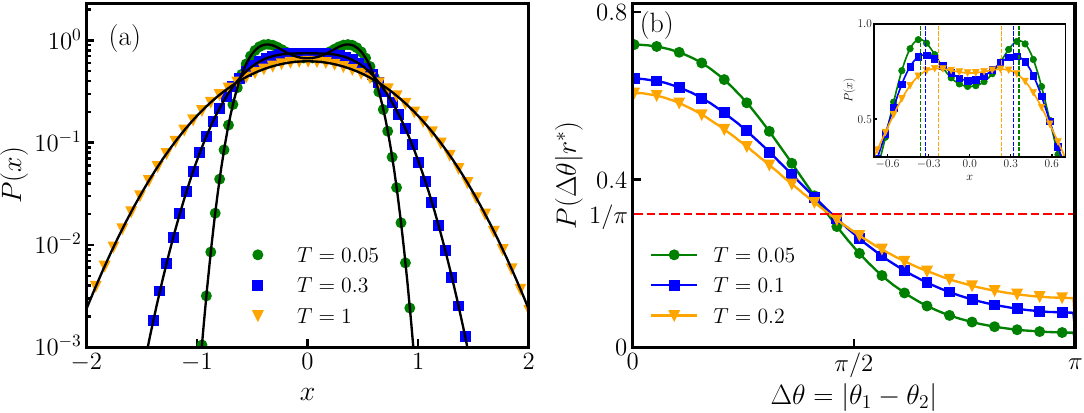}
    \caption{Strong-coupling regime: Stationary distribution of (a) the $x$-component of the separation ${\bm r}$, and (b) the relative orientation $P(\Delta\theta|r^*)$ for different temperatures $T$ obtained from numerical simulations with $v_1=4$, $v_2=1$, $D_1=D_2=0.01$, and $k=4$. In the inset of (b), $P(x)$ is shown along with the $r^*$ using dashed vertical lines. The black solid lines in (a) and red dashed line in (b) correspond to Eq.~\eqref{marg_dist_1} and uniform distribution $U(0,\pi)$, respectively.}
    \label{fig:strong_coupling_px_ang_dist}
\end{figure}

To compute the marginal distribution $P(x)$, we follow a similar approach as used in the previous section. From Eqs.~\eqref{x_comp_r} and \eqref{eq:Pr_th_int}, we can write the probability density function of $x$, for fixed $(\theta_1, \theta_2)$, as,
\begin{align}
    P(x|\theta_1,\theta_2)=\int_{-\infty}^{\infty} d w_x \, G(w_x) \delta\left[x-\left(\frac{v_1}{2 k}\cos{\theta_1}-\frac{v_2}{2 k}\cos{\theta_2}+w_x\right)\right].\label{cond_prob}
\end{align}
To proceed further, it is convenient to compute the Fourier transform of $P(x|\theta_1,\theta_2)$ with respect to $x$. This leads to, 
\begin{align}
    \tilde P(\lambda| \theta_1, \theta_2) \equiv \int_{-\infty}^{\infty} dx \, e^{i\lambda x} P(x|\theta_1,\theta_2) =  \frac{e^{-\frac{T \lambda^2}{2k}}}{2 \pi}\cos{\left(\frac{\lambda}{2 k}[v_1 \cos{\theta_1-v_2\cos{\theta_2}]}{}\right)}.
\end{align}
Integrating over $\theta_1$ and $\theta_2$, we have the characteristic function of the $x$-marginal distribution,
\begin{align}
    \tilde P(\lambda) \equiv \la e^{i \lambda x} \ra  = J_0\left(\frac{v_1 \lambda}{2k} \right) J_0\left(\frac{v_2 \lambda }{2k}\right) \exp \left(- \frac{T \lambda^2}{2 k}\right),\label{ft_11}
\end{align}
where $J_\nu(z)$ denotes the $\nu$-th order Bessel function of the first kind \cite{DLMF}. Inverse Fourier transform of $\tilde{P}(\lambda)$ gives the marginal distribution $P(x)=\int_{-\infty}^{\infty}d\lambda \cos{(\lambda x)} \tilde{P}(\lambda)/(2 \pi)$. To compute the integral over $\lambda$, we use the following identity of the Bessel function,
\begin{align}
J_0(a\,\lambda) J_0(b\,\lambda) = \sum_{\ell=0}^{\infty} \frac{(-1)^{\ell} a^{2\ell}}{(\ell!)^2} {}_2F_1 \left(-\ell,-\ell;1; \frac{b^2}{a^2} \right) \left(\frac{\lambda }{2}\right)^{2\ell}.\label{besel}
\end{align}
Substituting the above equation in Eq.~\eqref{ft_11} and integrating term by term, we get $P(x)$ as an infinite series sum,
\begin{align}
 P(x) = \sqrt{\frac k {2 \pi^2 T}} \sum_{l=0}^\infty  \frac{\Gamma(l+ \frac 12)}{(l!)^2} \left(- \frac{v_1^2}{8 kT} \right)^l {}_2F_1 \left(-l,-l;1; \frac{v_2^2}{v_1^2}\right) {}_1F_1\left(l+ \frac 12; \frac 12;-\frac{kx^2}{2T}\right),\label{marg_dist_1} 
\end{align}
were $\Gamma(z)$ denotes the Gamma function. \textcolor{black}{It can be explicitly shown that the stationary distribution $P(x)$ is symmetric under the exchange of $v_1$ and $v_2$ using the properties of the Hypergeometric function.} The marginal distribution $P(x)$ can be obtained up to arbitrary accuracy by numerically evaluating the series term by term. This is illustrated in Fig.~\ref{fig:strong_coupling_px_ang_dist}(a), which compares the $P(x)$ computed from Eq.~\eref{marg_dist_1} along with the same obtained from numerical simulations.  For temperatures below the threshold value $T^*$, 
$P(x)$ is bimodal with peaks at $x=\pm r^*$, signalling the presence of repulsion. At high temperatures $T> T^*$, the repulsion disappears, leading to a single-peaked distribution.

\subsubsection{Statistics of relative orientation:} In the absence of thermal noise, the minimum distance between the two particles is accompanied by the internal orientation of the particles being nearly parallel~\cite{sarkar2024emergent}. It is interesting to investigate the behaviour of the relative orientation of the two particles in the presence of thermal noise. From numerical simulations, we measure the conditional 
distribution $P(\Delta \theta|r^*)$ of the relative orientation $\Delta \theta = |\theta_1 - \theta_2|$ when the particles are separated by $r^*$, peaks of the stationary marginal $x$-distribution $P(x)$. Figure~\ref{fig:strong_coupling_px_ang_dist}(b) shows a plot of $P(\Delta \theta|r^*)$ for different temperatures $T$. Clearly, the most probable value of $\Delta \theta$ is close to zero, indicating that the orientations of the particles are nearly parallel when they are separated by their most probable distance. However, the distribution widens as the temperature is increased.

\subsubsection{The special case $v_1=v_2$:} 
\begin{figure}[t]
    \centering
    \includegraphics[width=0.9\linewidth]{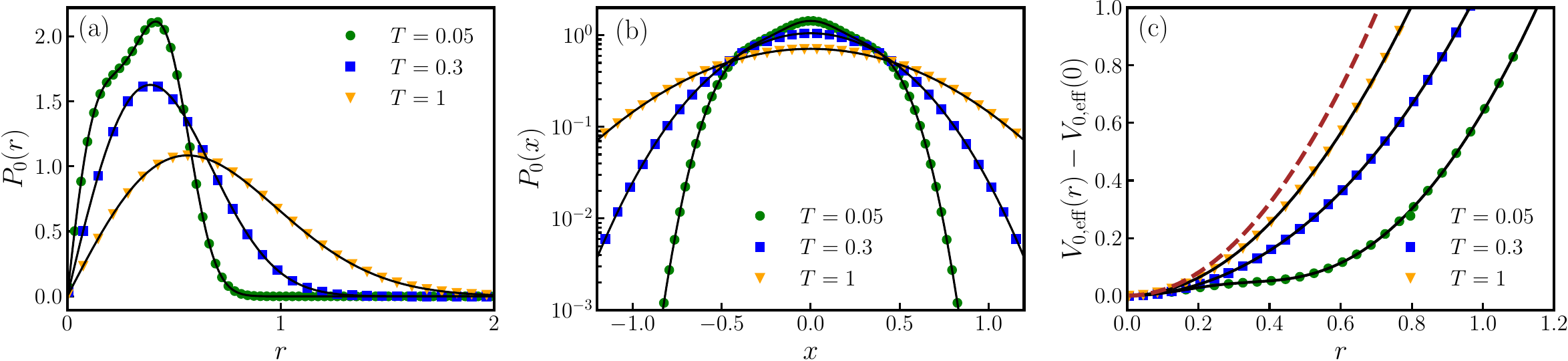}
    \caption{Strong-coupling regime with $v_1=v_2$: Stationary distribution of (a) the radial separation $r$, (b) the $x$-component of separation ${\bm r}$, and (c) the corresponding effective potential for different values of temperature $T$ obtained from numerical simulations with $v_1=v_2=v_0=2$, $D_1=D_2=0.01$, and $k=4$. The black solid lines in panel (a), (b), and (c) correspond to Eqs.~\eqref{radial_dist_strong_2}, \eqref{marg_dist_2} and \eqref{eff_pot2} respectively. The brown dashed line in (c) corresponds to the harmonic potential $k r^2/2$ with $k=4$. 
    }
    \label{fig:strong_coupling_eq_v}
\end{figure}

For the special case  $v_1=v_2=v_0$, the radial distribution in Eq.~\eqref{radial_dist_strong} reduces to,
\begin{align}
    P_0(r)
    &=\frac{kr}{T}\sum_{\ell=0}^\infty\frac{((2l)!)}{(\ell!)^3}\left(-\frac{v_0^2}{8 k T}\right)^\ell{}_1F_1\left(\ell+1,1,-\frac{k r^2}{2 T}\right).\label{radial_dist_strong_2}
\end{align}
This can be checked easily by using the following identity for the Hypergeometric function~\cite{DLMF},
\begin{align}
    {}_2F_1(a,b,c,1)=\frac{\Gamma(c)\Gamma(c-a-b)}{\Gamma(c-a)\Gamma(c-b)},~~\text{for}~~\mathrm{Re}(c-a-b)>0.\label{hyp_identity}
\end{align}
Although the sum over $\ell$ can not be performed to obtain a closed form for $P_0(r)$, even in this, it can be evaluated numerically in a straightforward manner.  Identity~\eref{hyp_identity} can also be used in Eq.~\eref{marg_dist_1} to obtain the $x$-marginal distribution in this case,
\begin{align}
 P_0(x) = \sqrt{\frac k {2 \pi T}} \sum_{l=0}^\infty \frac{\left((2l)!\right)^2}{(l!)^5}   \left(- \frac{v_0^2}{32 kT} \right)^l  {}_1F_1\left(l+\frac{1}{2};\frac{1}{2};-\frac{kx^2}{2T}\right).\label{marg_dist_2} 
\end{align}
Figure~\ref{fig:strong_coupling_eq_v}(a) and (b) illustrate the radial and $x$-marginal distributions for this special case. It should be noted that in this case, $P_0(x)$ is a single-peaked distribution for all $T$, indicating that the effective interaction is always attractive in the absence of speed diversity. This also becomes apparent from the nature of the effective potential, which is plotted in Fig.~\ref{fig:strong_coupling_eq_v}(c). Clearly, $V_{0,\text{eff}}(r)$ attains its minimum value at $r=0$, indicating an effective attraction between the particles for $v_1=v_2$. That can also be seen from the behaviour of $V_{0,\text{eff}}(r)$ near $r=0$---from Eq.~\eqref{b_eq} we get,
\begin{align}
    B = 8kT \left[1+ z \left(\frac {I_1(z)}{I_0(z)}-1\right)\right],\quad \text{where}\quad z=\frac{v_0^2}{4kT},
\end{align}
which is always positive. 

\textcolor{black}{To quantitatively characterize the effect of speed diversity on the separation fluctuations, it is useful to compute the `statistical distance' between the stationary radial and $x$-marginal distributions for the $v_1 \ne v_2$ case from the corresponding $v_1=v_2$ case. This can be done by measuring the Kullback–Leibler (KL) divergence of $P(r)$ and $P(x)$  with respect to $P_0(r)$ and $P_0(x)$, respectively. Indeed, we find that the KL divergence between the distributions increase as $v_1 - v_2$ is increased [see \ref{kl_div} for details].}

\subsection{Moderate-coupling regime}
The moderate-coupling regime emerges when the rotational diffusion coefficient of one of the particles, say $D_2$, is much larger than both $D_1$ and $k$. As shown in Fig.~\ref{fig:surface}(b), the qualitative behaviour of $P(x,y)$ in this regime is similar to the strong-coupling regime, with the particles most likely to be at a finite distance from each other. In this section, we compute the separation distribution in this regime analytically.

It is well known that for large $D_2$ and $v_2$, the effective noise correlations for the second particle in Eq.~\eqref{active_noise_correlation} reduces to,
\begin{align}
\la \cos \theta_2(t) \cos \theta_2(t') \ra &= \la \sin \theta_2(t) \sin \theta_2(t') \ra \simeq 2 T' \delta(t-t') \quad \mathrm{with,}\quad T'=\frac{v_2^2}{2 D_2},\label{eff_temp}   
\end{align}
and the typical motion of the ABP with high rotational diffusion constant resembles a Brownian motion with effective temperature $T$~\cite{bechinger_active_matter,abp_2D}.  Thus, in this regime, the time-evolution of the separation vector [see Eq.~\eref{eom_rel_dist}] can be effectively described by, 
\begin{align}
    \dot{\bm r}(t)
    =-2 k {\bm r}(t)+v_1 {\hat {\bm n}}_1(t)+\sqrt{2T_\mathrm{eff}}\,{\bm \xi}_\mathrm{eff}(t),\text{~and~}   
    \dot{\theta_{1}}(t) =\sqrt{2D_{1}}\, \eta_{1}(t),\label{moderate_coupling_eom}
\end{align}
where $\{ {\bm \xi}_\mathrm{eff} =(\xi_\mathrm{eff}^x, \xi_\mathrm{eff}^y) \}$ denote independent white noises  with zero mean and correlations given by
\begin{align}
\la \xi_\mathrm{eff}^\alpha(t) \xi_\mathrm{eff}^\beta (t') \ra = \delta_{\alpha \beta}\delta{(t-t')}.
\end{align}
and the effective temperature $T_\mathrm{eff}=2T+T'$.

It should be noted that Eq.~\eref{moderate_coupling_eom} is identical to the Langevin equation describing the motion of a single ABP in a harmonic trap of stiffness $2k$ and in contact with a thermal reservoir at temperature $T_\mathrm{eff}$. The corresponding stationary distribution of the radial coordinate $r$ has been computed by Malakar et al. in Ref.~\cite{abp_in_trap}. In the following, we use that result to obtain the radial and $x$-marginal distribution of the separation vector in this regime.

\subsubsection{Radial distribution:} 
\begin{figure}[t]
    \centering
    \includegraphics[width=0.7\linewidth]{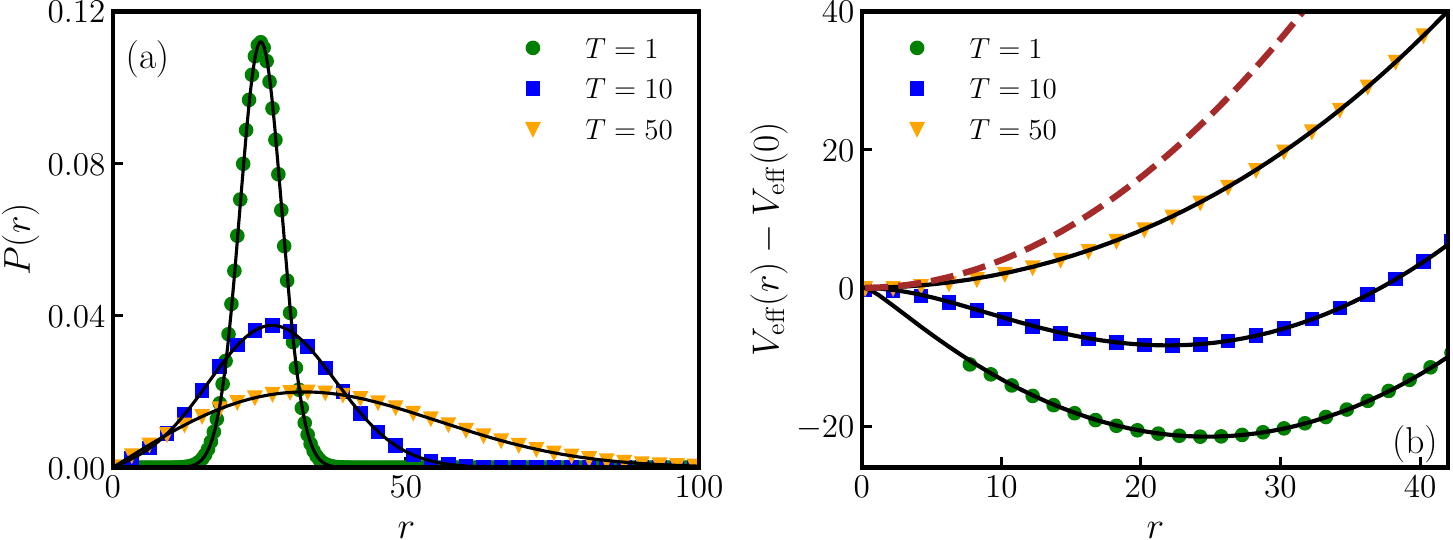}
    \caption{Moderate-coupling regime: (a) Stationary distribution of the radial separation $r$ and (b) the corresponding effective potentials for different values of temperature $T$ obtained from numerical simulations with parameters $v_1=4$, $v_2=1$, $D_1=10^{-3}$, $D_2=10$ and $k=0.08$. The black solid lines in (a) and (b) correspond to Eqs.~\eqref{moderate_r_dist} and \eqref{eff_pot_mod}, respectively. The brown dashed line in (b) corresponds to the underlying harmonic potential $k r^2/2$.}
    \label{fig:moderate_coupling_pr_eff_v}
\end{figure}

The moderate-coupling is characterized by $D_1 \ll k \ll D_2$. To the leading order, the radial distribution in this regime can be obtained by taking the $D_1 \to 0$ limit in the radial distribution computed in Ref.~\cite{abp_in_trap} [see Eq.~(C19) therein]. For our case, this translates to,
\begin{align}
P(r) = \frac{2k r}{T_\text{eff}} \exp{\left[- \frac{v_1^2+4 k^2 r^2}{4k T_\text{eff}} \right]}  I_0\left(\frac{v_1 r}{T_\text{eff}}\right),\label{moderate_r_dist}
\end{align} 
where, $I_0(z)$ denotes the zero-th order modified Bessel function of the first kind~\cite{DLMF}. Figure~\ref{fig:moderate_coupling_pr_eff_v}(a) compares this theoretical prediction with the $P(r)$ obtained from numerical simulations for different values of the temperature $T$, which show an excellent agreement.  Similar to the strong-coupling regime, the radial distribution $P(r)$ gets broader as the temperature is increased.

To understand the effective interaction between the two ABPs in the moderate-coupling regime, we calculate the effective potential defined in Sec.~\ref{strong_couple_pr}. Using 
Eq.~\eqref{eff_pot2} along with \eqref{moderate_r_dist}, we get the effective potential in the moderate-coupling regime,
\begin{align}
    V_\mathrm{eff}(r)-V_\mathrm{eff}(0)=\frac{T}{T_\mathrm{eff}}k r^2-T\log I_0\left(\frac{v_1 r}{T_\mathrm{eff}}\right).\label{eff_pot_mod}
\end{align}
Figure~\ref{fig:moderate_coupling_pr_eff_v}(b) shows the behavior of the effective potential, which is qualitatively similar to the strong-coupling regime. At small temperatures, $V_\text{eff}(r)$ has a minimum at some $r^*>0$, indicating the emergence of an effective repulsion between the two particles. The repulsion disappears as the temperature is increased beyond a threshold value $T^*$. As before, this threshold temperature can be estimated by looking at the small $r$ behaviour of $V_\text{eff}(r)$. Near $r=0$, 
\begin{align}
     V_\mathrm{eff}(r)-V_\mathrm{eff}(0) = \frac{T}{2 T_\text{eff}^2}\left(4 k T_\text{eff}-v_1^2\right) r^2 + \mathcal{O}[r^4].
\end{align}
Thus, the effective interaction remains repulsive for $T< T^*$, and,
\begin{align}
    T^*=\frac{v_1^2}{8 k}-\frac{T'}{2},\label{trans_temp_moderate}
\end{align}
where $T'$ is defined in Eq.~\eqref{eff_temp}. It should be emphasized that for $v_1^2<4kT'$, the effective interaction between the two ABPs is always attractive, even for $T=0$.

It should be noted here that the physical origin of the emergent repulsion in the moderate-coupling regime is different than that in the strong-coupling regime.  As mentioned already, the effective dynamics of the relative coordinate in the moderate-coupling regime is similar to a single active particle in an external harmonic potential, and the effective repulsion emerges from the fact that active particles tend to stay away from the minimum of the potential~\cite{abp_in_trap, solon2015active, abp_trap_1, abp_trap_2, abp_trap_3, abp_trap_4, PhysRevE.99.012145}. On the other hand, the repulsion in the strong-coupling regime originates solely due to the difference in the self-propulsion speeds of the particles.

\subsubsection{Marginal $x$-distribution:}\label{marginal_dist_moderate}

\begin{figure}[t]
    \centering
    \includegraphics[width=0.7\linewidth]{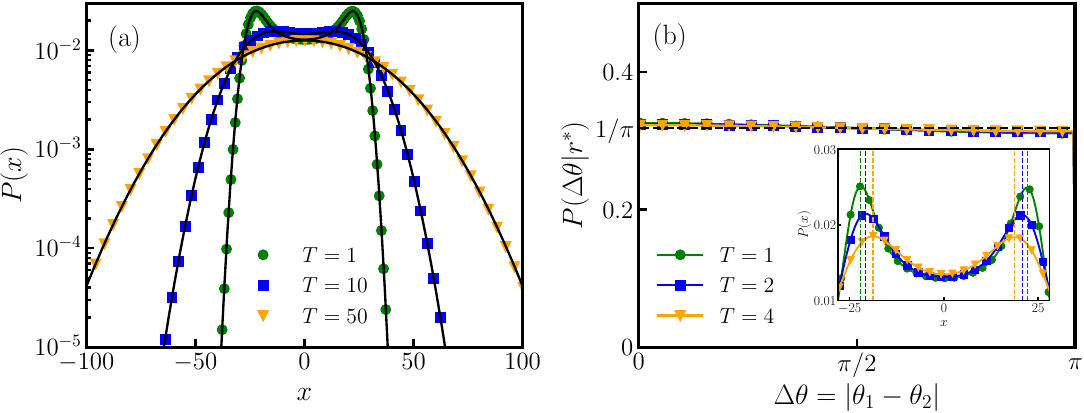}
    \caption{Moderate-coupling regime: Stationary distribution of (a) the $x$-component of separation ${\bm r}$, and (b) the relative orientation $\Delta\theta$ for given $r=r^*$ for different temperatures $T$, obtained from numerical simulations with parameters $v_1=4$, $v_2=1$, $D_1=10^{-3}$, $D_2=10$ and $k=0.08$. In the inset of (b), $P(x)$ is shown along with the $r^*$ using dashed vertical lines. The black solid lines in (a) correspond to Eq.~\eqref{moderate_x_dist} and the black dashed line in (b) corresponds to the uniform distribution $U(0,\pi)$.}
    \label{fig:moderate_coupling}
\end{figure}

We now turn to the $x$-marginal distribution in the moderate-coupling regime. This can be obtained in a straightforward manner from the radial distribution, exploiting the rotational symmetry of the system, 
\begin{align}
    P(x) = \int_{-\infty}^{\infty} \frac{dy}{2 \pi \, \sqrt{x^2+y^2}}\, P(r=\sqrt{x^2 + y^2}),
\end{align}
where $P(r)$ is given by Eq.~\eqref{moderate_r_dist}. Note that, for notational convenience, we have used the same letter $P$ to denote both radial and $x$-marginal distributions. The integral over $y$ can be performed by expanding $I_0(z)$ in series. This leads to,
\begin{align}
P(x)
&=\sqrt{\frac{T_\mathrm{eff}}{k\, \pi}}\exp{\left(-\frac{v_1^2+4 k^2 x^2}{4 k T_\mathrm{eff}}\right)} \sum_{\ell=0}^\infty \frac{(2 \ell)!{(k T_\mathrm{eff})^\ell}}{2^{4\ell}(\ell!)^3}
{}_{1}{F}_1\left(-\ell,\frac{1}{2}-\ell,-\frac{k}{T_\mathrm{eff}}x^2\right),\label{moderate_x_dist}
\end{align}
where ${}_1F_1(a,b,c)$ denotes the Kummer or Confluent hypergeometric function~\cite{DLMF}. The marginal distribution $P(x)$ obtained from numerical simulations for different values of $T$ is shown in Fig.~\ref{fig:moderate_coupling}(a) along with the analytical prediction given by Eq.~\eqref{moderate_x_dist}. Similar to the strong-coupling regime, the marginal distribution is bimodal with two peaks at $x=\pm r^*$ at temperatures less than the threshold temperature $T^*$ and becomes unimodal at temperatures higher than $T^*$ given by Eq.~\eqref{trans_temp_moderate}.

\subsubsection{Statistics of relative orientation:} 

It is also useful to measure the relative orientation distribution in the moderate-coupling regime. Figure~\ref{fig:moderate_coupling}(b) shows a plot of the conditional distribution $P(\Delta \theta|r^*)$ for different values of $T<T^*$, where $r^*$ denotes the peaks of the stationary marginal
$x$-distribution $P(x)$, obtained from numerical simulations. Clearly, $P(\Delta \theta|r^*)$ is almost uniform in the range $[0, \pi]$, unlike the strong-coupling case. This is due to the fact that, in the moderate-coupling regime the orientation vector $\bm{\hat n}_2$ changes very rapidly, and hence, even though $\bm{\hat n}_1$ evolves slowly, the relative orientation $|\theta_1 - \theta_2|$ attains a uniform distribution. This also signals the difference of the effective repulsion observed in the moderate-coupling regime with that in the strong-coupling regime.

\subsection{Weak-coupling regime}

\begin{figure}[t]
    \centering
    \includegraphics[width=0.9\linewidth]{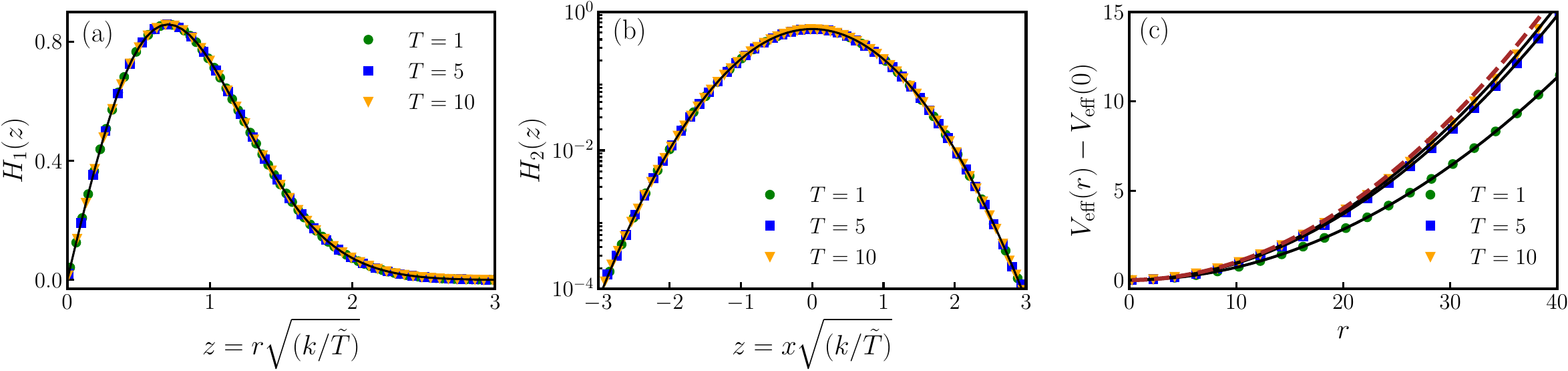}
    \caption{ Weak-coupling regime: Plot of the stationary distribution of (a) radial separation $r$, (b) the $x$-component of ${\bm r}$ for different values of the temperature $T$. (c) The corresponding effective potentials are shown in (c). Symbols indicate the data obtained from numerical simulations, while the black solid lines in (a), (b), and (c) correspond to Eq.~\eqref{eq_scaled} and \eqref{eff_pot_we},  respectively. The brown dashed line in (c) corresponds to the harmonic potential $k r^2/2$. Here we have taken $v_1=4$, $v_2=1$, $D_1=10$, $D_2=20$ and $k=0.02$.  }
    \label{fig:weak_coupling}
\end{figure}

The weak-coupling regime emerges when both the rotational diffusion constants $D_1$ and $D_2$ are much larger than the coupling strength $k$, i.e., when the activity time-scales of both the ABPs are much smaller than the relaxation time-scale in the trap. In this case, for large $v_1$ and $v_2$, the typical motions of both the ABPs are similar to passive Brownian particles with effective diffusion constant $v_1^2/(2D_1)$ and $v_2^2/(2D_2)$, respectively [see Eq.~\eqref{eff_temp}]. Consequently, the Langevin equation Eq.~\eqref{eom_rel_dist} in the weak-coupling regime can be rewritten as,
\begin{align}
    \dot{\bm r}(t)  
    = -2 k {\bm r}(t) + \sqrt{2\tilde T}~ {\bm \eta}_\mathrm{eff}(t), ~\text{with}~~ \tilde T = 2T+\frac{v_1^2}{ 2D_1}+\frac{v_2^2}{2D_2},\label{weak_coup_eom}
\end{align}
where $\{ {\bm \eta}_\mathrm{eff} =(\eta_\mathrm{eff}^x, \eta_\mathrm{eff}^y) \}$ denote independent white noises with zero mean and correlations $\la \eta_\mathrm{eff}^\alpha(t) \eta_\mathrm{eff}^\beta (t') \ra = \delta_{\alpha \beta}\delta{(t-t')}.$ The above equation describes the two-dimensional motion of a passive particle in a harmonic trap of strength $2k$ at \textcolor{black}{an effective temperature $\tilde T$, that depends on the activity parameters $v_1,v_2$ and $D_1, D_2$}. Consequently, $P(\bm{r})$ is nothing but a two-dimensional Gaussian distribution. Hence, it is straightforward to show that the stationary radial and $x$-marginal distributions must have the scaling form,
\begin{align}
 P(r) = \sqrt{\frac k {\tilde T}} \, H_1 \left(r \sqrt{\frac k {\tilde T}} \right), \quad \text{and}, P(x) = \sqrt{\frac k {\tilde T}} \, H_2 \left(x \sqrt{\frac k {\tilde T}} \right),\label{dist_weak}
\end{align}
where the scaling functions are given by,
\begin{align}
 H_1(z) = 2z e^{- z^2}, \quad \text{and}, \quad H_2(z) = \frac 1{\sqrt{\pi}} e^{-z^2}. \label{eq_scaled}
\end{align}
Figure~\ref{fig:weak_coupling}(a) and (b) show scaled plots of $P(r)$ and $P(x)$, respectively, for different temperatures. 

As expected, in the weak-coupling regime, the two particles are most likely to be near each other. This also becomes apparent from the effective potential,
\begin{align}
    V_\mathrm{eff}(r)-V_\mathrm{eff}(0)
    =\frac{T}{\tilde T}k r^2,\label{eff_pot_we}
\end{align}
which remains harmonic, albeit with a modified coupling constant. At high temperatures, the effective potential coincides with the actual underlying potential $kr^2/2$, as $T\to \infty$, $T/\tilde T\to1/2$ [see Eq.~\eqref{weak_coup_eom}]. Figure~\ref{fig:weak_coupling}(c) illustrates the behaviour of the effective potential for different values of $T$.

\textcolor{black}{It should be noted that Eqs.~\eqref{dist_weak} describe the typical fluctuations of the separation vector ${\bm r}$ in this regime---signatures of activity are expected to be observed in the large deviations of $r$ and $x$.}

\section{Motion of the centroid}\label{sec:cm}

To comprehensively understand the behaviour of the binary system, we also characterize the motion of the centroid ${\bm R}(t)$, which evolves according to the Langevin equation \eref{eom_cm}. Clearly, the centroid motion is unaffected by the coupling strength of the coupling. At late-times, i.e., times larger than the active time-scales, the centroid is expected to show a diffusive motion with typically Gaussian position fluctuation. On the other hand, signatures of activity are expected at short-times, which we characterize in this section.

\subsection{Position moments} 

First, we compute the mean and variance of the position of the centroid. The formal solution of Eq.~\eref{eom_cm} reads,
\begin{align}
    R_x(t) &=\int_0^t ds \left[\frac{v_1}{2} \cos{\theta_1(s)}+\frac{v_2}{2} \cos{\theta_2(s)}+\sqrt{T}\zeta^x(s) \right], \label{cm_int_x} \\ 
    R_y(t) &=\int_0^t ds \left[\frac{v_1}{2} \sin{\theta_1(s)}+\frac{v_2}{2} \sin{\theta_2(s)}+\sqrt{T}\zeta^y(s) \right], \label{cm_int_y}  
\end{align}
where $R_x$ and $R_y$ denote the $x$ and $y$ components of the centroid position ${\bm R}$. Clearly, $\la R_x(t) \ra = \la R_y(t) \ra =0$ at all times. The second moments of $R_x(t)$ and $R_y(t)$ can be computed using Eq.~\eref{active_noise_correlation} and are given by,
\begin{align}
\la R_x^2 (t) \ra =\la R_y^2 (t) \ra = Tt +  \frac{v_{1}^2}{4 D_1^2}\left(D_1 t+e^{-D_1t}-1\right)+  \frac{v_{2}^2}{4 D_2^2}\left(D_2 t+e^{-D_2t}-1\right).\label{moment_centeroid}
\end{align}

As expected, at late times $t \gg (D_1^{-1}, D_2^{-1})$,
\begin{align}
 \la R^2(t) \ra \equiv   \la R_x^2 (t) \ra + \la R_y^2 (t) \ra \simeq \left[2T + \frac {v_1^2}{2 D_1} + \frac {v_2^2}{2 D_2}\right] t
\end{align}
which indicates the diffusive nature of the centroid motion. On the other hand, at short-times $t \ll (D_1^{-1}, D_2^{-1})$, 
\begin{align}
    \la R_x^2(t) \ra \simeq T t+\left(\frac{v_1^2+v_2^2}{8}\right)t^2+\mathcal{O}[t^3].
\end{align}
The presence of the activity adds a ballistic component to the short-time dynamics of the centroid, similar to single active particles~\cite{bechinger_active_matter,howse_motile_colloid,martens2012probability,zheng2013non}. Fig.~\ref{fig:cm_dist}(a) compares the theoretical prediction Eq.~\eqref{moment_centeroid} with the second moment of the centroid obtained from numerical simulations for different bath temperatures $T$, which shows an excellent agreement.

\subsection{Short-time marginal distribution}

\textcolor{black}{Short-time position fluctuations of active particles show prominent signatures of activity, including anisotropy and strong non-Gaussianity, as observed for single particles~\cite{abp_2D, short_t_dist, Das_2023} as well as interacting systems~\cite{Prakash_2024,D4SM00350K,Singh_2021}.} In this section, we investigate the signatures of activity in the short-time distribution of the centroid position and how it differs from that of a single particle. 

\begin{figure}[t]
    \centering
    \includegraphics[width=0.7\linewidth]{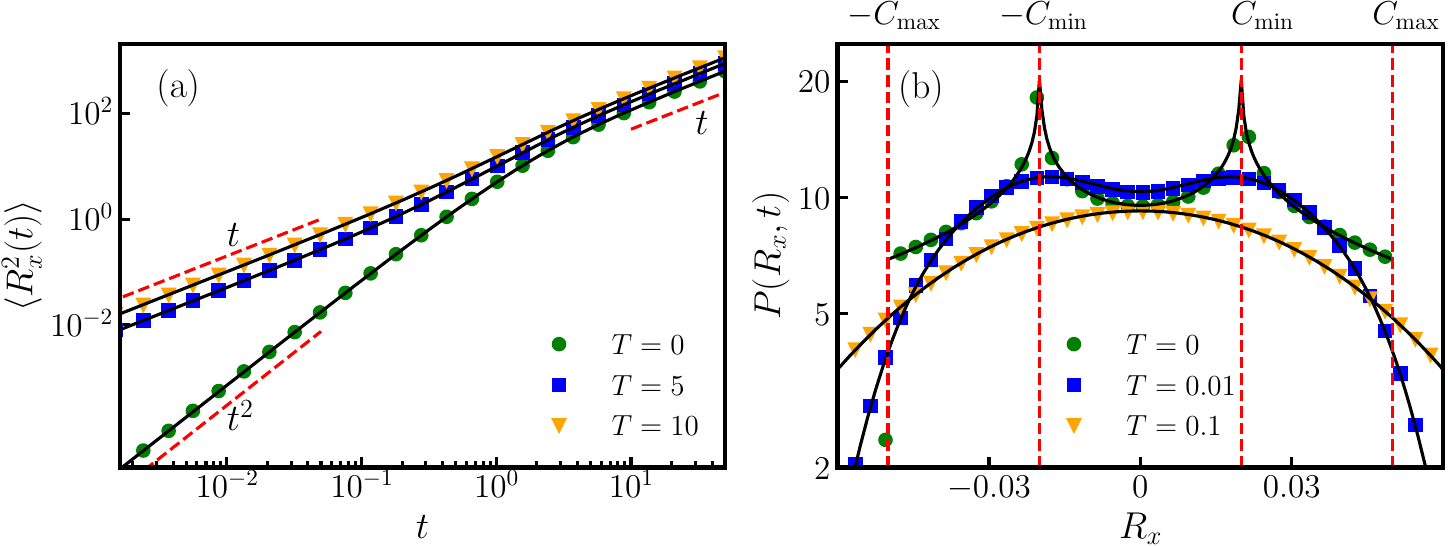}
    \caption{Motion of the centroid: Plot of $\la R_x^2(t) \ra$ as a function of time  (a)  and short-time distribution $P(R_x,t)$ at $t=0.01$ (b) for different temperatures $T$, obtained from numerical simulations. For both (a) and (b), we have taken $k=1$, $D_1=1$, $D_2=4$, $v_1=7$, $v_2=3$. Black solid lines in (a) and (b) correspond to Eq.~\eqref{moment_centeroid} and Eq.~\eqref{short_dist_analytic}, respectively.}
    \label{fig:cm_dist}
\end{figure}
We adopt a trajectory-based approach and focus on $P(R_x,t)$, the $x$-marginal distribution of the centroid position. To this end, it is convenient to write $\theta_i=\psi_i+\phi_i$ where $\psi_i$ denotes the initial orientation of the $i$-th particle, drawn independently from the uniform distribution in $[0,2\pi]$, and $\phi_i$ undergoes a Brownian motion [see Eq.~\eref{eq:model_phi}],
\begin{align}
    \dot{\phi}_i(t)&=\sqrt{2 D_i}\,\eta_i(t).\label{angular_dissusion}
\end{align}
At short-times, i.e., for $t\ll (D_1^{-1}, D_2^{-1})$, $\phi_i$ is also small, and we can approximate $\sin \phi_{i} \approx \phi_i $ and $\cos \phi_{i} \approx 1$.  Thus, in this regime, Equation~\eqref{cm_int_x} reduces to, 
\begin{align}
R_x(t) ={\cal A} t - {\cal B}_1 \int_0^t ds~ \phi_1(s) - {\cal B}_2 \int_0^t ds~ \phi_2(s) + \sqrt{T}\, \int_0^t ds~\zeta^x(s)\label{pRx_short_terms}
\end{align} 
where, 
\begin{align}
    \mathcal{A}=\frac{1}{2}(v_1 \cos{\psi_1}+v_2 \cos{\psi_2}),\quad \text{and}~ \mathcal{B}_i=\frac{v_i}2 \sin{\psi_i}.
\end{align}
Each of the three terms on the right-hand side of Eq.~\eqref{pRx_short_terms} corresponds to an integral of a Gaussian process. In particular, $\int_0^t ds \, \phi_i(s)$ corresponds to a random acceleration process~\cite{rap_1,rap_2,rap_3} with variance  $\frac23 D_i\,t^3$. Thus, for given $(\psi_1, \psi_2)$, the distribution of $R_x(t)$ is also a Gaussian with mean $\mathcal{A}t$ and variance $\frac23t^3(D_1\mathcal{B}_1^2+D_2\mathcal{B}_2^2)+tT$, 
\begin{align}
    {\cal P}(R_x,t| \psi_1, \psi_2) =  \frac{\exp{\left[-\frac{(R_x-\mathcal{A}t)^2}{ 2 tT+\frac43t^3(\mathcal{B}_1^2\,D_1+\mathcal{B}_2^2 D_2)}\right]}}{\sqrt{ \pi  \left(2tT+\frac43t^3(\mathcal{B}_1^2 D_1+\mathcal{B}_2^2 D_2)\right)}}.
\end{align}
The short-time probability distribution of $R_x$ can now be obtained by integrating over all the uniform distributions of $(\psi_1, \psi_2)$,
\begin{align}
    P(R_x,t)=\int_0^{2\pi}\frac{d \psi_2}{2 \pi} \int_0^{2\pi} \frac{d \psi_1}{2 \pi} \frac{\exp{\left[-\frac{(R_x-\mathcal{A}t)^2}{ 2 tT+\frac43t^3(\mathcal{B}_1^2\,D_1+\mathcal{B}_2^2 D_2)}\right]}}{\sqrt{ \pi  \left(2tT+\frac43t^3(\mathcal{B}_1^2 D_1+\mathcal{B}_2^2 D_2)\right)}}.\label{short_dist_analytic}
\end{align}
Although it is hard to evaluate the integrals analytically, the distribution $P(R_x,t)$ can be estimated accurately by performing the integrals numerically.  Figure~\ref{fig:cm_dist}(b) illustrated the behavior of $P(R_x,t)$ for different values of temperature $T$. For $T=0$, i.e., in the absence of thermal bath, the motion of the centroid is bounded in the region  $C_\mathrm{min}\leq R \equiv |{\bm R}(t)|\leq C_\mathrm{max}$ [see Eq.~\eqref{bound_cm}]. As a result, the marginal distribution $P(R_x,t)$ has divergence at $C_\mathrm{min}$ and cut-off at $C_\mathrm{max}$. 
For $T>0$, the distribution $P(R_x, t)$ gets smoother, and at large temperatures, it eventually becomes Gaussian with zero mean and variance $tT$.

\section{Conclusion}\label{sec:concl}

In this paper, we study the behaviour of two harmonically coupled active Brownian particles in the presence of thermal noise. In particular, we characterize the nonequilibrium stationary state of the separation of the two particles and the short-time dynamics of the centroid. We show that the stationary separation statistics shows very different behaviour depending on the relative strengths of the two active time-scales and the relaxation time-scale corresponding to the coupling, which we characterize analytically. 

At low temperatures, both the strong-coupling and moderate-coupling regimes are characterized by the emergence of an effective short-range repulsion between the two particles. The effective repulsion disappears as the temperature is increased beyond a threshold value $T^*$, which we obtain analytically. We further characterize the emergent repulsion by computing the effective potential $V_\text{eff}(r)$, which shows a minimum at a finite $r^* >0$. However, the physical origin of the repulsion in these two regimes are different. In the strong-coupling regime, the repulsion emerges solely due to the difference in the self-propulsion speeds of the two particles. In fact, when the propulsion speeds of the two particles are same, they feel an effective attraction. On the other hand, in the moderate-coupling regime, the effective repulsion is similar to the high probability of a single active particle being away from the minimum of an external potential.

With the rapid progress in \textcolor{black}{robotics~\cite{PRXLife.2.033007, paramanick2024programming,paramanick2025spontaneous,li2019particle, PhysRevE.98.052606}, and colloidal physics~\cite{magnetic_colloid, Buttinoni_2012,halder2025interplay},} it is possible to investigate active matter systems in controlled ways. Therefore, we expect that the emergence of the short-range repulsion between harmonically coupled active particles can be observed in an experimental setup. Furthermore, recent studies suggest that diversity in propulsion speed and attractive interaction oppose cluster formation in active systems \textcolor{black}{in both lattice~\cite{D1SM01009C, SciPostPhys.14.6.165, Ray_2024} and continuum frameworks~\cite{PhysRevE.91.042310,C3SM52469H,PhysRevE.108.064613}.} It is worth exploring whether the minimal theoretical framework considered here can account for these observations.

\section*{Data availability statement}
The data are available from the authors upon reasonable request.

\section*{Acknowledgements}
R.S. acknowledges the CSIR grant no. 09/0575(11358)/2021-EMR-I. U.B. acknowledges the support from the Anusandhan National Research Foundation (ANRF), India, under a MATRICS grant [No. MTR/2023/000392].

\section*{Conflict of interest}
There are no conflicts to declare.

\appendix
\section{Statistical separation of stationary distributions in the strong-coupling regime}\label{kl_div}
\textcolor{black}
{
\begin{figure}[t]
    \centering
    \includegraphics[width=0.7\linewidth]{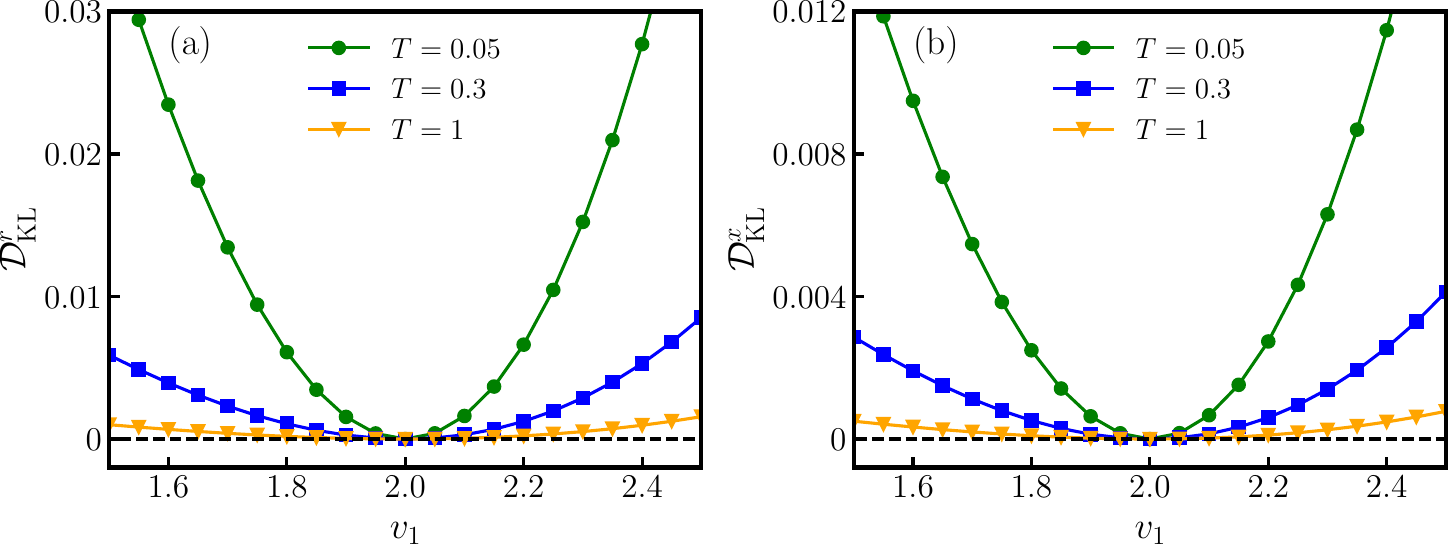}
    \caption{Plot of Kullback–Leibler divergence for (a) stationary radial distribution and (b) $x$-marginal distribution as functions of $v_1$ in the strong-coupling regime for different temperatures $T$, obtained from numerical integration of Eq.~\eqref{kl}. Here we have taken $v_2=2$, $v_0=2$ and $k=4$.}
    \label{fig:kl_div}
\end{figure}
}
\textcolor{black}
{To measure the \textit{statistical distance} of the stationary separation distribution in the $v_1 \ne v_2$ from that in the $v_1=v_2$ case, it is useful to calculate the Kullback–Leibler (KL) divergence~\cite{kullback1951information}. The KL divergences for the radial and $x$-marginal distributions in the strong-coupling regime are given by,
\begin{align}
    \mathcal{D}^r_\mathrm{KL}=\int_0^{\infty} dr \, P(r)\, \log{\left(\frac{P(r)}{P_0(r)}\right)},\quad \mathrm{and}\quad \mathcal{D}^x_\mathrm{KL}=\int_{-\infty}^{\infty} dx \, P(x)\, \log{\left(\frac{P(x)}{P_0(x)}\right)}, \label{kl}
\end{align}
where $P(r)$, $P_0(r)$, $P(x)$ and $P_0(x)$ is given by Eqs.~\eqref{radial_dist_strong}, \eqref{radial_dist_strong_2}, \eqref{marg_dist_1} and \eqref{marg_dist_2}, respectively. 
}

\textcolor{black}{The KL divergences corresponding to the radial and $x$-marginal distributions can be obtained by numerically performing the above integrals using the explicit forms of $P(r)$, $P_0(r)$, $P(x)$ and $P_0(x)$. Figure~\ref{fig:kl_div} shows plots of  $\mathcal{D}^r_\mathrm{KL}$ and $\mathcal{D}^x_\mathrm{KL}$ as functions of $v_1$, for different values of temperature. Clearly, the `distance' between the distriutions vanish for $v_1=v_2$,  while it increases as $v_1 - v_2$ is increased. As expected, the difference between the distributions are more pronounced at low temperatures. }

\section*{Bibliography}
\bibliographystyle{iopart-num}
\bibliography{ref}
\end{document}